\documentstyle[preprint,aps,tighten,bezier,floats]{revtex}

\def\be{\begin{equation}}
\def\ee{\end{equation}}
\def\bea{\begin{eqnarray}}
\def\eea{\end{eqnarray}}
\def\beas{\begin{eqnarray*}}
\def\eeas{\end{eqnarray*}}
\def\no{\nonumber}
\def\a{\alpha}

\def\G{\Gamma}
\def\d{\delta}

\def\m{\mu}
\def\n{\nu}
\def\LLL{\Lambda}
\def\lll{\lambda}

\def\O{\Omega}
\def\p{\phi}
\def\vp{\varphi}

\def\ps{\psi}
\def\r{\rho}

\def\t{\widetilde}

\def\bra{\langle}
\def\ket{\rangle}
\newcommand{\lbra}{\left\langle}
\newcommand{\rket}{\right\rangle}

\def\eq{\equiv}

\def\pa{\partial}
\def\ra{\rightarrow}

\def\intdx{\int\! d^4\!x\,}

\newcommand{\np}[1]{Nucl.\ Phys.\ {\bf {#1}}}
\newcommand{\plt}[1]{Phys.\ Lett.\ {\bf {#1}}}

\newcommand{\pr}[1]{Phys.\ Rev.\ {\bf {#1}}}
\newcommand{\prlt}[1]{Phys.\ Rev.\ Lett.\ {\bf {#1}}}
\newcommand{\ijmp}[1]{Int.\ J.\ Mod.\ Phys.\ {\bf {#1}}}

\newcommand{\anp}[1]{Ann.\ Phys. (N. Y.)\ {\bf {#1}}}
\newcommand{\cmp}[1]{Comm.\ Math.\ Phys.\ {\bf {#1}}}

\newcommand{\jph}[1]{J.\ Phys.\ {\bf {#1}}}
\def\ps{\psi}
\newcommand{\lo}{{\Lambda_0}}
\newcommand{\dph}{{\cal D}\phi}
\newcommand{\dvl}[2]{{\cal D}_{#1,#2}^{\LLL}}

\newcommand{\lfilo}{S_{int}^{(f)\lo}}
\newcommand{\lfil}{S_{int}^{(f)\Lambda}}
\newcommand{\litlo}{S_{\scriptstyle{\rm tot}}^{\lo}}
\newcommand{\litl}{S_{\scriptstyle{\rm tot}}^{\Lambda}}
\newcommand{\lilo}{S_{int}^{\lo}}
\newcommand{\lil}{S_{int}^{\Lambda}}

\newcommand{\lilon}{S_{int}^{\lo;N}}
\newcommand{\lilonf}{S^{\lo;N}}
\newcommand{\lilog}[1]{S_{int}^{\lo;#1}}

\newcommand{\liln}{S_{int}^{\Lambda;N}}
\newcommand{\lilg}[1]{S_{int}^{\Lambda;#1}}

\newcommand{\ltilog}[1]{\widetilde S_{int}^{\lo;#1}}

\newcommand{\wtl}{W_{\scriptstyle{\rm tot}}^\Lambda}
\newcommand{\wl}{W^\Lambda}
\newcommand{\gl}{\G^\Lambda}
\newcommand{\gil}{\G_{int}^\Lambda}
\newcommand{\lfl}[2]{{\cal G}_{#1,#2}^{(f)\LLL}}
\newcommand{\lflo}[2]{{\cal G}_{#1,#2}^{(f)\lo}}
\newcommand{\lfo}[2]{{\cal G}_{#1,#2}^{(f)\LLL=0}}

\newcommand{\lvl}[2]{{\cal G}_{#1,#2}^{\LLL}}
\newcommand{\lvlo}[2]{{\cal G}_{#1,#2}^{\lo}}
\newcommand{\lvlon}[2]{{\cal G}_{#1,#2}^{\lo;N}}

\newcommand{\lvo}[2]{{\cal G}_{#1,#2}^{\LLL=0}}

\newcommand{\lvon}[2]{{\cal G}_{#1,#2}^{\LLL=0;N}}

\newcommand{\lvll}[2]{{\cal G}_{#1,#2}^{\lambda}}

\newcommand{\lvln}[2]{{\cal G}_{#1,#2}^{\LLL;N}}

\newcommand{\ltog}[3]{\widetilde{\cal G}_{#2,#3}^{\LLL=0;#1}}

\newcommand{\ltln}[2]{\widetilde{\cal G}_{#1,#2}^{\LLL;N}}
\newcommand{\ltlg}[3]{\widetilde{\cal G}_{#2,#3}^{\LLL;#1}}

\newcommand{\ltpln}[2]{\widetilde{\cal G'}_{#1,#2}^{\LLL;N}}

\newcommand{\lbl}[2]{\bar{\cal G}_{#1,#2}^{(f)\LLL}}
\newcommand{\lbo}[2]{\bar{\cal G}_{#1,#2}^{(f)\LLL=0}}

\def\dlo{{P^{\lo}}}
\newcommand{\dl}{{P^{\Lambda}}}
\newcommand{\dllo}{{P_{\Lambda}^{\lo}}}
\newcommand{\dmlo}{{P_m^{\lo}}}
\newcommand{\dml}{{P_m^{\Lambda}}}

\newcommand{\dMlo}{{P_M^{\lo}}}
\newcommand{\dMl}{{P_M^{\Lambda}}}

\newcommand{\pal}{\pa_\Lambda}

\newcommand{\plog}[2]{{\rm Plog}\left(\frac{#1}{#2}\right)}
\newcommand{\norm}[2]{\left\|{#1}\right\|_{#2}}
\newcommand{\norml}[1]{\left\|{#1}\right\|_{(2\LLL,\m)}}
\newcommand{\normn}[1]{\left\|{#1}\right\|_{(2\LLL,\m,M)}}
\newcommand{\intdp}{\int\frac{d^4p}{(2\pi)^4}}
\newcommand{\intdq}{\int\frac{d^4q}{(2\pi)^4}}

\newcommand{\mono}[2]{M_{#1,\{#2\}}}
\newcommand{\caln}{{\cal N}_{\{w\}}}

\begin{document}
\preprint{hep-th/9810056
\hspace{-33.2mm}\raisebox{2.4ex}{KIAS-P98026}
\hspace{-25.7mm}\raisebox{4.8ex}{SNUTP 98-100}}
\title{Wilson Renormalization Group and Continuum Effective Field
Theories\footnote{Lecture presented at The 11th Summer School and Symposium 
on Nuclear Physics (NuSS'98), ``Effective Theories of Matter'', Seoul 
National University, June 23--26, 1998, Korea}}
\author{Chanju Kim\thanks{e-mail: cjkim@kias.re.kr}}
\address{Center for Theoretical Physics\\
         Seoul National University Seoul, 151-742, Korea\\
         and\\
         Korea Institute for Advanced Study\footnote{present address}\\
         207-43 Chungryangri-dong Dongdaemun-ku, Seoul, 130-012, Korea}
\maketitle
\begin{abstract}
This is an elementary introduction to Wilson renormalization group and 
continuum effective field theories. We first review the idea of Wilsonian 
effective theory and derive the flow equation in a form that allows multiple
insertion of operators in Green functions. Then, based on this formalism, we
prove decoupling and heavy-mass factorization theorems, and discuss how 
the continuum effective field theory is formulated in this approach.
\end{abstract}
\newpage
\section{Introduction}
The purpose of this lecture is twofold. First, we will give an elementary
introduction to Wilson renormalization group in field theories developed 
recently. Then based on this formalism, we discuss some basic aspects of 
continuum effective field theories.

The concept of effective field theories has played an important role in modern
theoretical physics and it acquires its natural physical
interpretation in the Wilson renormalization group formalism \cite{wilson}.
In the latter, one integrates out the high frequency modes scanned by a
cutoff and then considers lowering the cutoff.
This generates the renormalization group flow, and
the differential equation governing this flow is called
the exact renormalization group equation or the flow equation.
After Wilson's formulation, this flow equation has been applied to wide area
of physics, from condensed matter physics to particle physics
\cite{wilsonkogut}, especially for nonperturbative problems which are 
difficult to treat in the other approaches. It has also greatly enhanced the
understanding of renormalization in quantum field theory. 

In spite of the understanding of renormalization with 
Wilsonian approach, the precise connection to the
conventional approach has, however, not been explicitly given until 
Polchinski was able to prove the perturbative renormalizability 
of a renormalizable quantum field theory within
this framework \cite{polchinski}, taking the $\p^4$-theory as his model 
field theory. His method was remarkably simple and could diffuse all
difficulties of the conventional approach. 
More recently, direct physical meaning of Wilsonian effective action was
given in the framework of conventional field theory that it is nothing but
the generating functional of Green functions with an {\it infrared} 
cutoff \cite{keller}. After this important observation, Wilsonian effective
action and its renormalization group equation became a powerful
tool to study wide range of nuclear and high-energy physics problems 
\cite{warr}-\cite{aoki}. 

If we take seriously the above point of view on effective field theory and
renormalization, it should be natural also to have a simple connection to the
conventional formulation of effective theory. Here, one of the most basic 
results is
the decoupling theorem \cite{applequist} which states that, in a generic
renormalizable quantum field theory with heavy particles of mass $M$,
heavy-particle effects decouple from low-energy light-particle physics 
except for renormalization of couplings involving light fields and 
corrections of order $1/M$. Thus the resulting renormalizable effective 
field theory describes low energy physics to the accuracy $1/M$. 
In other words, to the zeroth order in $1/M$ there are no observable effects
due to the existence of virtual heavy particles. If one wishes to
understand the low-energy manifestations of heavy particles, then irrelevant
(nonrenormalizable) terms should be considered to incorporate higher order
effects in $1/M$. Investigation of this issue shows that
the virtual heavy particle effects can be isolated via a set
of effective local vertices with calculable couplings. It implies that
low-energy light-particle physics can be described by a suitable local
effective field theory when combined with appropriate calculation rules to
deal with irrelevant vertices. This heavy-mass factorization was proved to
the lowest in $1/M$ adapting Zimmermann's algebraic
identities \cite{zimmermann} in the BPHZ formulation \cite{clee}, 
and, recently, to all orders in perturbation theory and to to any given 
order in powers of $1/M$ using the flow equation \cite{ckim3} (see also Refs.\
\cite{girardello,ball2}).
Actually the proof is almost as simple as the renormalizability proof itself 
and the whole scheme allows a natural physical interpretation from 
the viewpoint of the renormalization group flow as we will see later.

There is, however, an issue to clarify regarding the difference
between Wilsonian effective theories and effective theories in continuum.
We have three widely separated scales here---light particle mass
$m$, heavy particle mass $M$, and the ultraviolet (UV) cutoff $\lo$, with
$m\ll M\ll\lo$. And our hope is to find a low-energy effective theory with
the UV cutoff $\lo$ which involves only the
light field of mass $m$ and describes low-energy light-particle physics
accurately up to the order $1/M^{N}$ ($N$ is any fixed integer).
Then the effective theory must include a finite number of irrelevant terms
(restricted by the dimensional argument). Here arises a problem.
In the original Wilson's view, the cutoff scale of the effective
theory may well be identified with the heavy mass scale $M$, above
which it is no more effective. The presence of irrelevant terms in the
effective Lagrangian at scale $M$ is then very natural as discussed above
and there is no need to worry about their presence in particular. The
``natural scale'' of those terms will be around $M$, i.e., they are
order one at scale $M$, and among them we may choose to keep explicitly some
minimal number of irrelevant terms\footnote{For
example, the continuum version of Symanzik's improved action \cite{symanzik}
in
lattice theory have been discussed in \cite{keller4} by adding suitable
irrelevant terms in Lagrangian.} in our effective Lagrangian for the accuracy
of order $1/M^{N}$.
But, in the conventional discussion of quantum field
theory, the UV cutoff $\lo$ is supposed to go to infinity eventually.
So, to connect Wilson's view with this conventional formulation, we may
suppose scaling up the cutoff of the above Wilsonian effective Lagrangian
from $M$ to
$\lo$. It will then generate infinitely many irrelevant terms which are
unnaturally large. Also, during the scaling, all of them
get mixed together and
so we are forced to work with the Lagrangian consisting of infinitely many
terms all the time. Any kind of truncation for the bare Lagrangian to some
finite number of terms would yield divergences in physical quantities as
$\lo\ra\infty$, because the unnaturally large coefficients would be amplified
by some positive power of $\lo/M$ as the cutoff is scaled down.
This is nothing but the statement of nonrenormalizability in the
language of renormalization group flow.
To avoid this problem we need to deal with irrelevant terms
carefully, i.e., give suitable rules to obtain unambiguous finite results with
only a finite number of terms included in the bare Lagrangian. It is
achieved through the modification of the flow equation when irrelevant
vertices are inserted, in the more-or-less same way as one treats
the renormalization of composite operators
and their normal products \cite{zimmermann} in the
flow equation approach \cite{keller2}. We will discuss this procedure 
later.

The plan of the paper is as follows. In section 2, 
starting from the generating functional, we identify
the Wilsonian effective action and give a rather lengthy derivation of
the Wilson renormalization group equation\cite{zinn,morris}. 
This derivation will show clearly how 
the Wilsonian effective action is interpreted as a generating 
functional of Green functions. Then we derive the flow equation in 
a form applicable to the case that multiple insertion of 
composite operators is allowed in Green
functions. In section 3, after a review on perturbative renormalizability,
we show decoupling and heavy-mass factorization taking as our model a
scalar theory involving two real scalar fields of masses $m$ and $M$ with
$m\ll M$. Then we discuss how 
the continuum effective field theory is formulated to any
desired order of accuracy in the flow equation approach. We conclude in
section 4.

\section{The flow equation}

For simplicity let us consider a theory of single scalar field in four 
Euclidean dimensions with a momentum space cutoff $\lo$. 
Generalization to several fields is
straightforward. The bare action is written as
\be
S^{\lo}[\phi]=\frac12\lbra\p,\dlo^{-1}\p\rket+\lilo[\p]\,,
\ee
where $\dlo$ is the free-particle cutoff propagator and $\bra f,g\ket$ is
defined by the momentum-space integral of $f$ and $g$,
\be
\bra f,g\ket\equiv\intdp f(p)g(-p)\,,
\ee
and $\lilo[\p]$ represents the interaction part of the bare action. 
For later use, we will also allow the insertions
of some additional local vertices or certain composite operators in 
Green functions of the theory. To account for this, we define 
$\litlo[\p]$ as a formal power series in $\a$,
which has $\lilo[\p]$ as its zeroth order term,\footnote{More generally,
one can consider power series of many parameters $\a_1,\ldots,\a_k$.  
Equation (\ref{litlo}) and subsequent discussion can 
accommodate this case with 
the interpretation that $\a=(\a_1,\ldots,\a_k)$, $N$ is 
a multi-index $N=(N_1,\ldots,N_k)$, $|N|=\sum_{i=1}^kN_i$,
$N!= N_1!\cdots N_k!$, and $\a^N=\a_1^{N_1}\cdots\a_k^{N_k}$.}
viz.,
\be \label{litlo}
\litlo[\p]
  =\sum_{|N|\ge0}\frac{\a^N}{N!}\lilon[\p]\,.
     \quad (\lilo[\p]\eq\lilog{0}[\p])
\ee
Also, for later conveniences, we will denote
\be
S_C^\lo[\p]=\sum_{|N|\ge1}\frac{\a^N}{N!}\lilon[\p]
        \qquad\left(=\litlo[\p]-\lilo[\p]\right)\,.
\ee
$\lilon$'s may be regarded as additional local vertices appended to the 
original Lagrangian $\lilo$ or as composite operators in which one is 
interested.
For example, if one wishes to consider a single or twice 
insertion of a composite operator ${\cal O}(x)$, one may consider
\bea \label{comp}
\lilog{1}&=&\intdx \chi(x){\cal O}(x)\,,\no\\
\lilog{2}&=&\intdx d^4y\chi(x)\chi(y){\cal O}'(x,y)\,,
\eea
where $\chi(x)$ is the source for ${\cal O}(x)$ and ${\cal O}'(x,y)$ is a 
suitable counterterm for the product of operators ${\cal O}(x){\cal O}(y)$, 
which can be determined
through renormalization conditions and the flow equation derived below
\cite{keller2}. In this case, $\litlo$ is equal to the original
interaction part of the action plus the composite-operator source terms.

The generating functional, with the insertion of the operator
$e^{S_C^\lo[\p]}$, is
\be\label{zj}
Z[J]=\int\dph e^{-\frac12\bra\p,\dlo^{-1}\p\ket-\litlo[\p]+\bra J,\p\ket}.
\ee
Following the idea of Wilson and Polchinski \cite{wilson,polchinski}, 
we wish to integrate out the high-momentum components of $\p$
and reduce the cutoff $\lo$ to a lower scale $\LLL$. 
For this, we divide the propagator $\dlo$ into the high-frequency 
part $\dllo$ and the low-frequency part $\dl$, the borderline being set by
momentum\footnote{There
will be infinite ways in doing the separation; choosing one way of 
separation may be considered as choosing a ``renormalization
scheme'' in the Wilson renormalization group approach, 
and physical quantities are independent of such choices. 
It is not necessary to specify a particular scheme at this stage.}
$p=\LLL$:
\be
\dlo=\dl+\dllo\,,
\ee
Then it is not difficult to show that the generating functional may be 
written in terms of two fields $\p_H$ and $\p_L$ rather than $\p$ alone as
\be\label{zj2}
Z[J]=\int\dph_H\dph_L\,
 e^{-\frac12\bra\p_H,\dllo^{-1}\p_H\ket-\frac12\bra\p_L,\dl^{-1}\p_L\ket
    -\litlo[\p_H+\p_L]+\bra J,\p_H+\p_L\ket}\,,
\ee
up to a multiplicative factor. (The equivalence of Eq.\ (\ref{zj}) and Eq.\
(\ref{zj2}) may be seen if one substitutes $\p_L=\p-\p_H$ into Eq.\
(\ref{zj2}) and performs the integral over $\p_H$ which is Gaussian.)
Since the field $\p_L(\p_H)$ has $\dl(\dllo)$ as its propagator, only the
low-(high-)frequency modes of $\p_L(\p_H)$ will now propagate 
effectively. The integral over $\p_H$ may be performed to obtain
\cite{morris}
\be
Z[J]=\int\dph_Le^{-\frac12\bra\p_L,\dl^{-1}\p_L\ket}
                     e^{-\wtl[\p_L,J]}\,,
\ee
where $\wtl$ is given by
\begin{equation} \label{wtl}
e^{-\wtl[\p_L,J]}\equiv\int\dph_He^{-\frac12\bra\p_H,\dllo^{-1}\p_H\ket
  -\litlo[\p_H+\p_L]+\bra J,\p_H+\p_L\ket}\,.
\end{equation}
Notice that $\wtl[0,J]$ is nothing but the 
generating functional of connected Green
functions (with the operator $e^{S_C^\lo[\p]}$ inserted) in the theory with 
{\it both\/} UV cutoff $\lo$ and IR cutoff $\LLL$.
Substituting $\p_H=\p-\p_L$ into Eq.\ (\ref{wtl}), we may write
\begin{eqnarray}\label{lil}
e^{-\wtl[\p_L,J]}&=&e^{-\frac12\bra\p_L,\dllo^{-1}\p_L\ket}\int\dph
 e^{-\frac12\bra\p,\dllo^{-1}\p\ket-\litlo[\p]+\bra J+\dllo^{-1}\p_L,\p\ket}
                                        \no\\
&=&e^{\frac12\bra J,\dllo J\ket+\bra J,\p_L\ket}
   e^{-\frac12\bra J+\dllo^{-1}\p_L,\dllo(J+\dllo^{-1}\p_L)\ket}\no\\
& &\times e^{-\litlo[\frac{\d}{\d J}]}
   e^{\frac12\bra J+\dllo^{-1}\p_L,\dllo(J+\dllo^{-1}\p_L)\ket}\no\\
&=&e^{-\wtl[0,0]}
   e^{\frac12\bra J,\dllo J\ket+\bra J,\p_L\ket-\litl[\dllo J+\p_L]}\,,
\end{eqnarray}
for some $\litl[\p]$ satisfying $\litl[0]=0$. (Here we factored out 
the field independent part as $e^{-\wtl[0,0]}$ which goes to 1 as 
$\LLL\ra\lo$.) Therefore, the generating functional can be written, up to a
multiplicative factor, as
\begin{equation}
Z[J]=\int\dph e^{-\frac12\bra\p,\dl^{-1}\p\ket-\litl[\dllo J+\p]
                 +\bra J,\p\ket+\frac12\bra J,\dllo J\ket}\,,
\end{equation}
where we have replaced $\p_L$ by $\p$.
Suppose that $J(p)=0$ for $p>\LLL$ so that $J$ couples to the 
low-frequency modes only. 
Then, since $\dllo$ has only high-frequency modes, all $J$'s drop out 
from the expression
except for $\bra J,\p\ket$ and $\litl$ coincides with the Wilsonian
effective action \cite{wilson}. However, we do not have to insist on such a
restriction for $J(p)$; for general $J(p)$, Eq.\ (\ref{lil}) connects 
directly $\litl$ with Green functions as will be discussed shortly.

Now we derive the flow equation \cite{polchinski} satisfied by $\litl$.
Differentiating Eq.~(\ref{wtl}) with respect to $\LLL$ while choosing
$\p_L=0$, we have the result
\be \label{flowW}
\pal e^{-\wtl[0,J]}=-\frac12
    \lbra\frac{\d}{\d J},\pal\dllo^{-1}\frac{\d}{\d J}\rket e^{-\wtl[0,J]}\,,
\ee
which, on using Eq.~(\ref{lil}) for $e^{-\wtl[0,J]}$, gives the flow 
equation,
\be \label{flow}
\pal\litl[\p]=\left.-\frac12\intdp\pal\dl(p)
       \left[\frac{\d^2\litl}{\d\p(p)\d\p(-p)}
      -\frac{\d\litl}{\d\p(p)}\frac{\d\litl}{\d\p(-p)}\right]
      \right|_{\rm field\mbox{-}dep.\ part}\,.
\ee
Equation (\ref{flow}) is integrable; the formal solution with the 
appropriate boundary condition satisfied at $\LLL=\lo$ is \cite{keller}
\be  \label{sol}
e^{-\litl[\p]-I^\LLL}
   =e^{-\frac12\lbra\frac{\d}{\d\p},\dllo\frac{\d}{\d\p}\rket}
              e^{\litlo[\p]}\,,
\ee
where $I^\LLL$ is supposed to collect precisely all $\phi$-independent 
pieces from the right hand side so that we may have $\litl[0]=0$ as we 
required in (\ref{lil}). Equation (\ref{sol}), together with 
Eq.~(\ref{lil}), reproduces a well-known expression for
the generating functional $e^{\wtl}$. In fact, $\litl$ bears a simple 
connection with physical amplitudes. To see this, notice that we have, 
from Eq.~(\ref{lil}),
\be \label{imp}
\wtl[0,J]-\wtl[0,0]=-\frac12\bra J,\dllo J\ket+\litl[\dllo J]\,,
\ee
and hence
\be
\left.\frac{\d^n\litl[\p]}{\d\p(p_1)\cdots\d\p(p_n)}\right|_{\p=0}
=\left.\prod_{i=1}^n\dllo^{-1}(p_i)
 \frac{\d^n\wtl}{\d J(p_1)\cdots\d J(p_n)}\right|_{J=0}\,,
     \hspace{0.5cm}n>2\,.
\ee
Therefore, we arrive at a very interesting result. Namely, Wilsonian 
effective action $\litl$ with {\it UV} cutoff $\LLL$ can also be 
interpreted as the generating functional of amputated connected Green 
functions (with the operator $e^{S_C^\lo[\p]}$ inserted) with {\it IR} 
cutoffs $\LLL$. 
Physical Green functions are then obtained in the limit $\LLL\ra0$
\cite{keller,bonini,morris}. Actually, the reason behind this result
is rather simple. The generating functional of a field theory is obtained
by integrating out {\it all} modes, all informations of the 
theory being encoded in the dependence on the external source coupled to 
the field. If a cutoff $\LLL$ is introduced and integration is performed 
only over high frequency modes, then we will get the generating 
functional with IR cutoff $\LLL$ because low frequency modes remain 
unintegrated. On the other hand, from the point of view
of low frequency modes, the resulting functional gives, by definition, 
the Wilsonian effective action. Equation (\ref{imp}) precisely expresses
this relation.

Now that Wilsonian effective action is essentially the same as the
generating functional of connected Green functions, it is also interesting
to perform a Legendre transformation to obtain the flow equation for the
generating functional of one-particle irreducible Green functions with
IR cutoff $\LLL$. As usual, we define
\bea
\gl[\vp] &\equiv & \wl[0,J] + \bra J,\vp \ket \no\\
         &=& \gil[\vp] + \frac12 \bra \vp, \dllo^{-1} \vp\ket\,,
\eea
with
\be
\frac{\d\gl}{\d\vp} = J\,,\quad \frac{\d\wl}{\d J} = -\vp\,.
\ee
Then from Eq.~(\ref{flowW}), we obtain \cite{wetterich}
\bea
\pal\gil &=& \frac12 \intdp\pal \dllo^{-1}(p)
             \frac{\d^2\wl}{\d J(p)\d J(-p)} \no\\
  &=& \frac12 \intdp\pal \dllo^{-1}(p)
\left[ \frac{\d^2\gil}{\d\vp(p)\d\vp(-p)} + \dllo^{-1}\right]^{-1}\,.
 \label{flowG}
\eea
It is often more convenient to work with 1PI quantities than connected 
ones, especially for the purpose of practical calculations 
\cite{wetterich2,morris}. However, we will not discuss it further.  See 
Refs.\ \cite{berges,bonini,morris} for details.
\vspace{3mm}

Let us go back to $\litl$. Expanding $\litl$ in powers of $\a$, 
we define $\liln$'s as the expansion coefficients, i.e.,
\be
\litl[\p]=\sum_{|N|\ge0}\frac{\a^N}{N!}\liln[\p]\,, 
      \quad\left(\lilg{0}[\p]\eq \lil[\p]\right)\,.
\ee
As we mentioned above, $\liln[\p]$ is the generating functional of amputated
connected Green functions with an insertion of an operator $O_N$ which is
identified as the $N$-th order coefficient of $e^{S_C^\lo[\p]}$, i.e.,
\be \label{oiexpansion}
\sum_{|N|\ge1}\frac{\alpha^N}{N!}O_N\eq e^{S_C^\lo}
 =\exp{\bigl(\sum_{|N|\ge1}\frac{\alpha^N}{N!}\lilon\bigr)}\,.
\ee
For example, $O_1=\lilog{1}$, $O_2=\lilog{2}+(\lilog{1})^2$,
$O_3=\lilog{3}+3\lilog{2}\lilog{1}+(\lilog{1})^3$ and so
on.\footnote{Expression for general $O_N$ is also available in \cite{ckim3}.}
At the zeroth order of $\a$, we obtain the flow equation for the effective
Lagrangian $\lil$ from Eq.~(\ref{flow}), which assumes an identical form as
Eq.~(\ref{flow}) other than the replacement $\litl\ra\lil$.
At the $N$-th order of $\a$, on the other hand, we have
\bea \label{flowk}
\pal\liln&\hspace{-2mm}=&\hspace{-2mm}
    -\frac12\intdp\pal\dl(p)\left[\frac{\d^2\liln}{\d\p(p)\d\p(-p)}
    -2\frac{\d\lil}{\d\p(p)}\frac{\d\liln}{\d\p(-p)}\right.\no\\
    &&\hspace{46mm}\left.\left.
    -\sum_{0<I<N}\pmatrix{N\cr I}\frac{\d S_{int}^{\LLL;I}}{\d\p(p)}
     \frac{\d S_{int}^{\LLL;N-I}}{\d\p(-p)} \right]
     \right|_{\rm field\mbox{-}dep.\ part}\,,\no\\
  &&
\eea
For a single insertion of operator through $S_{int}^{\lo;1}$, 
the last term of the right hand side in the second line of
Eq.\ (\ref{flowk}) vanishes and so we have a linear and homogeneous equation
in $S_{int}^{\LLL;1}$. This equation has been used in discussing the
renormalization of composite operators \cite{hughes,keller2}. 
Equation (\ref{flowk}) with multi index $N=(1,1)$ 
is useful in discussing the 
short-distance expansion of two composite operators \cite{keller2}. 
Notice that, if $N>1$, Eq.\ (\ref{flowk}) contains inhomogeneous terms which 
may look difficult to deal with. However, in spite of those terms,
it is still of the first order in $\liln$ while the homogeneous part 
remains exactly the same for all $N$. Therefore the general solution 
will be a sum of a particular solution and a solution to the 
homogeneous equation (which is just the $N=1$ equation for 
Green functions with a single insertion of an operator). 
Though we will not explicitly use this property here, it is important in 
considering Zimmermann's normal product 
and multiple insertion of local operators of which the formulation is crucial
in understanding the structure of effective theories \cite{ckim3}.

A few remarks are in order.

(i) As noted above, choice of the cutoff function in the propagator is quite
arbitrary. For example, the propagator $\dl$ of a particle with mass $m$ 
is given by
\be
\dl = \frac{R(\LLL,p)}{p^2+m^2}\,,
\ee
where $R(\LLL,p)$ is a cutoff function. If we choose sharp cutoff,
$R(\LLL,p) = \theta(\LLL-p)$. Sometimes it is more convenient to choose 
smooth cutoff such as $R(\LLL,p)= 1-e^{p^2/\LLL^2}$. More generally, 
even the separation of the action $S^\LLL$ into the free part and
the interaction part $\lil$ is arbitrary. One may write, for example,
\begin{equation} \label{wtl2}
e^{-\wtl[J]}\equiv\int\dph e^{-\frac12\bra\p,{\t{P}_\LLL^{\lo}\,}^{-1}\p\ket
  -S^\lo[\p]+\bra J,\p\ket}\,,
\end{equation}
where $S^\lo$ is the full bare action of the theory with UV cutoff $\lo$ and
\be
{\t{P}_\LLL^{\lo}\,}^{-1} \eq \dllo^{-1} - \dlo^{-1}
\ee
is the pure cutoff term added to the action. Then all the previous equations
are valid with $\lil$ and $\dllo$ replaced by $S^\LLL$ and 
${\t{P}_\LLL^{\lo}\,}^{-1}$, respectively. The corresponding flow equation 
for $\G^\LLL$ is the one used by Berges in this volume.

(ii) The flow equation (\ref{flow}) is too complicated to solve exactly. 
For practical purposes, 
it is therefore necessary to find suitable approximation methods.
An obvious way is to perform a derivative expansion
\be
 S^\LLL = \intdx \left[
    V(\vp,\LLL) + \frac12 (\pa_\m\vp)^2 Z(\vp,\LLL) + O(\pa^4)\right]\,.
\ee
Then the flow equation reduces to simple differential equations of 
coefficient functions $V(\vp,\LLL)$, $Z(\vp,\LLL)$ etc.\ and one can study
their properties by various means. There are some problems with 
approximation methods. As mentioned
above, physical  quantities should be independent of the choice of cutoff
functions. However, this scheme independence is lost after approximations.
Another problem is about reparametrization invariance: physics should not
depend on the reparametrization of fields in the partition function.
This is reflected in the flow equation in some complicated way \cite{repara}
and is broken in general by approximations. For the discussion of these
issues see, e.g., \cite{moreno,morris333}.

(iii) When the system has a gauge symmetry, integrating out high momentum
modes does not preserve gauge symmetry and this could be a potentially serious
problem. There are a few ways with which  this problem can be coped with. The
simplest way is to insist on using the gauge-symmetry-violated flow equation
derived here. Even if the gauge symmetry is not manifest for finite cutoff
$\LLL$, it is eventually restored at the physical point $\LLL=0$ up to
$O(1/\lo)$ \cite{warr,bonini,wetterich4,ellwanger}. 
Also, one may work in the background gauge 
in which the background gauge invariance may be maintained in the 
effective action \cite{wetterich4} though it may not necessarily mean
quantum gauge invariance \cite{dattanasio}.

\section{Construction of effective field theories}
As discussed in section 1, one of the most natural application of 
the flow equation would be to understand basic results 
of effective field theories in continuum. Here we discuss decoupling
and heavy-mass factorization, and so construct the continuum effective theory 
to any desired order of accuracy in flow equation approach. As a first step,
let us review Polchinski's proof of perturbative renormalizability
\cite{polchinski}.

\subsection{Perturbative Renormalizability}
A theory is said to be perturbatively renormalizable if Green functions of the
theory are bounded and converge to finite limits at each order of perturbation
as UV cutoff of the theory
goes to infinity. Now knowing that Wilsonian effective action gives physical
Green functions in the limit $\LLL\ra0$ and that it is controlled by the
flow equation, we may expect that the boundedness and convergence of Green
functions can be easily shown by estimating the flow equation. Indeed, this
line of argument can be implemented in a quite straightforward way. 
Since basically the same 
reasoning is repeatedly used throughout in discussing effective
theories, we first explain the method in rather detail. It consists of the
following steps:
\begin{enumerate}
\item[(i)] Write down the flow equation for vertex functions and 
identify boundary
conditions which follow from the form of the bare action (at $\LLL=\lo$) and
also from the renormalization conditions on Green functions (at $\LLL=0$).
\item[(ii)] Integrate the flow equation over $\LLL$ from the boundary 
at which boundary conditions are given (either at $\LLL=0$ or at $\LLL=\lo$)
and estimate the resulting expression.
\item[(iii)] Prove boundedness and convergence order 
by order using induction.
\end{enumerate}

Let us illustrate this with a $\p^4$-theory. The bare action with UV cutoff
$\lo$ reads
\be\label{esbare}
S^{\lo}[\p]=\frac12\lbra\p,\dmlo^{-1}\p\rket+\lilo\,,
\ee
where $\dmlo$ is the free-particle cutoff propagator of mass $m$ and $\lilo$
is the interaction part of the bare action as before. Explicitly,
\be  \label{cebare}
\lilo=\intdx\left[\frac12\r_1\p^2(x)+\frac12\r_2(\pa_\m\p(x))^2
        +\frac{1}{4!}\r_3\p^4(x)\right]\,.
\ee
In conventional way, the bare couplings $\r_a$ may be written as
\be \label{barec}
\r_1=Z_\p m_0^2-m^2,\qquad\r_2=Z_\p-1,\qquad\r_3=Z_\p^2 g^0,
\ee
where $m_0$, $Z_\p$, and $g^0$ are the bare mass, the wavefunction
renormalization, and the bare coupling of the real scalar field $\p$.
Propagator $\dmlo$ is defined by
\be \label{lightp}
\dmlo=\frac{R_m(\lo,p)}{p^2+m^2},
\ee
where $R_m(\LLL,p)$ is a smoothed variant of the sharp cutoff
function $\theta(\LLL-p)$. As we indicated in section 2, choosing $R_m$
corresponds to choosing a ``renormalization scheme'' and physical
quantities are not affected. Here, for convenience' sake, we choose
\be \label{cutoff}
R_m(\LLL,p)=\left[1-K\left((1+\LLL/m)^2\right)\right]
               K\left(\frac{p^2}{\LLL^2}\right)\,,
\ee
where $K(t)$ is a smooth function such that $K(t)\ra1$ (exponentially) 
as $t\ra1$ and $K(t)\ra0$ (also exponentially) as $t\ra 4$ (detailed form
of $K(t)$ is not important).

Now, the Wilsonian effective action $\lil$ is defined to satisfy the flow
equation (\ref{flow}). Expanding $\lil$ in powers of $\p$ in momentum space,
we
write
\bea \label{eexpansion}
&&\lil[\p]=\sum_{r=0}^\infty\sum_{n=1}^\infty g^r
          \int\prod_{j=1}^{2n-1}\frac{d^4p_j}{(2\pi)^4}\p(p_1)\cdots\p(p_{2n})
        \lvl{r}{2n}(p_1,\ldots,p_{2n-1})\,,\\
&&\hspace{80mm}
  {\textstyle\left(p_{2n}\eq-\sum_{j=1}^{2n-1}p_j\right)\,,}\no
\eea
where $g$ is the perturbation-expansion parameter which may be identified as
renormalized coupling constants.
Then as we noted above, at $\LLL=0$, vertex functions $\lvl{r}{2n}$ can be 
identified with amputated connected Green functions 
of the theory. Furthermore, we have $\lvl{r}{2n}=0$ if $n>r+1$ because just
the connected diagrams contribute to $\lvl{r}{2n}$. Therefore
the sum over $n$ in Eq.\ (\ref{eexpansion}) actually extend only over a
finitely many $n$ for given $r$.

If we insert the expansion (\ref{eexpansion}) into the flow equation
(\ref{flow}) we arrive at 
\bea\label{lvlflow}
\lefteqn{\pal\lvl{r}{2n}(p_1,\ldots,p_{2n-1})}\no\\
&&=-\pmatrix{2n+2\cr 2}\intdq \pal\dml(q)
 \lvl{r}{2(n+1)}(q,-q,p_1,\ldots,p_{2n-1})\no\\
&&\ +2\sum_{r=1}^{r-1}\sum_{l=1}^nl(n-l+1)\pal\dml(P)\no\\
 &&\hspace{20mm}\times\left[\lvl{r}{2l}(p_1,\ldots,p_{2l-1})
       \lvl{r-r'}{2(n-l+1)}(-P,p_{2l},\ldots,p_{2n-1})\right]_{symm}\,,\no\\
&&\hspace{40mm}{\textstyle\left(P=-\sum_{j=1}^{2n-1}p_j\right)}
\eea
where $[\cdots]_{symm}$
implies symmetrization with respect to momenta $p_1,\ldots,p_{2n}$. Denoting
$\pal\dml$ by a straight line, we can represent this equation by the
diagram shown in Fig.1. 

\begin{figure}
\setlength{\unitlength}{1mm}
\begin{picture}(0,0)(-20,-20)
\put(22.5,0){\makebox(0,0)[cc]{\LARGE =}}
\put(66,0){\makebox(0,0)[cc]{\LARGE +}}
\put(77.5,-2.5){\makebox(0,0)[cc]{$\displaystyle\sum_{l}$}}

\put(0,0){\circle{10}}
\put(-4.5,-2.25){\line(-2,-1){5.06}}
\put(4.5,-2.25){\line(2,-1){5.06}}
\put(-3.74,-3.74){\line(-1,-1){4}}
\put(-2.5,-7){\makebox(0,0)[cc]{$\cdot$}}
\put(0,-7.43){\makebox(0,0)[cc]{$\cdot$}}
\put(2.5,-7){\makebox(0,0)[cc]{$\cdot$}}
\put(4.7,-5.8){\makebox(0,0)[cc]{$\cdot$}}
\put(0,-12){\makebox(0,0)[cc]{$\sf 2n$}}
\put(-15,0){\makebox(0,0)[cc]{\LARGE$\partial_\Lambda$}}
\end{picture}
\begin{picture}(0,0)(-65,-20)
\put(0,0){\circle{10}}
\put(-4.5,-2.25){\line(-2,-1){5.06}}
\put(4.5,-2.25){\line(2,-1){5.06}}
\put(-3.74,-3.74){\line(-1,-1){4}}
\put(-2.5,-7){\makebox(0,0)[cc]{$\cdot$}}
\put(0,-7.43){\makebox(0,0)[cc]{$\cdot$}}
\put(2.5,-7){\makebox(0,0)[cc]{$\cdot$}}
\put(4.7,-5.8){\makebox(0,0)[cc]{$\cdot$}}
\put(0,-12){\makebox(0,0)[cc]{$\sf 2(n+1)$}}
\bezier{72}(-2,4.6)(-5,10)(0,11)
\bezier{72}(2,4.6)(5,10)(0,11)
\end{picture}
\begin{picture}(0,0)(-110,-20)
\put(0,0){\circle{10}}
\put(-4.5,-2.25){\line(-2,-1){5.06}}
\put(4.5,-2.25){\line(2,-1){5.06}}
\put(-3.74,-3.74){\line(-1,-1){4}}
\put(-2.5,-7){\makebox(0,0)[cc]{$\cdot$}}
\put(0,-7.43){\makebox(0,0)[cc]{$\cdot$}}
\put(2.5,-7){\makebox(0,0)[cc]{$\cdot$}}
\put(4.7,-5.8){\makebox(0,0)[cc]{$\cdot$}}
\put(0,-12){\makebox(0,0)[cc]{$\sf 2l$}}
\put(5,0){\line(1,0){16.5}}
\put(13.5,3){\makebox(0,0)[cc]{$\sf\partial_\Lambda P_m^\Lambda$}}
\end{picture}
\begin{picture}(140,40)(-135,-20)
\put(0,0){\circle{10}}
\put(-4.5,-2.25){\line(-2,-1){5.06}}
\put(4.5,-2.25){\line(2,-1){5.06}}
\put(-3.74,-3.74){\line(-1,-1){4}}
\put(-2.5,-7){\makebox(0,0)[cc]{$\cdot$}}
\put(0,-7.43){\makebox(0,0)[cc]{$\cdot$}}
\put(2.5,-7){\makebox(0,0)[cc]{$\cdot$}}
\put(4.7,-5.8){\makebox(0,0)[cc]{$\cdot$}}
\put(0,-12){\makebox(0,0)[cc]{$\sf 2(n-l+1)$}}
\end{picture}

\caption{Graphical representation of the flow equation (\ref{lvlflow}).}
\end{figure}
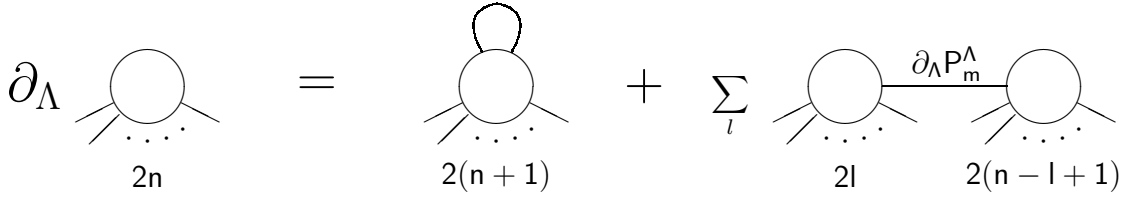

If suitable boundary conditions are given, the flow equation (\ref{lvlflow})
will produce the amputated connected Green functions of the theory,
$G_{r,2n}^{c}=\lvo{r}{2n}$.
First, we know that the bare action (\ref{cebare}) has a simple form:
at $\LLL=\lo$, 
\bea \label{vbc}
\pa_p^w\lvlo{r}{2n}=0,\hspace{45mm}&&2n+|w|>4\,,\no\\
\pa_p^w\lvlo{r}{2n}(0)\sim
 \mbox{bare couplings $\r_a$ in Eq.~(\ref{barec})},&&2n+|w|\le 4\,,
\eea
where $w$ is the multi index
$\{w\}\equiv\{w_1,\ldots,w_{2n-1}\}$ with $w_j=(w_{j1},\ldots,w_{j4})$
and $\pa_p^w\eq\pa_{p_1}^{w_1}\cdots\pa_{p_{2n-1}}^{w_{2n-1}}$,
$\pa_{p_i}^{w_i}\equiv\prod_{\m=1}^4\pa^{w_{i\m}}/(\pa p_{i\m})^{w_{i\m}}$.
We will also use the notations $|w|\equiv\sum_{i,\m}|w_{i\m}|$.
We do not know the value of bare couplings $\r_a$; they are fixed by
the renormalization conditions imposed on Green functions at $\LLL=0$ for
$2n+|w| \le 4$. For example, we may choose the conditions at zero momenta,
\be \label{effbc}
\lvo{r}{2}(0)=\pa_{p_\m}\pa_{p_\n}\lvo{r}{2}(0)=0,\qquad
\lvo{r}{4}(0)=\frac1{4!}\d_{r1}\,.
\ee

Now that the boundary conditions have been completely specified, we integrate
Eq.\ (\ref{lvlflow}) and estimate it. For that, we introduce a set
of norms \cite{keller2}, $\|\ \|_{(a,b)}$, where $a$ and $b$ are positive
real numbers:
\be \label{norm}
\norm{(\pa^z f)g}{(a,b)}\eq
 \max_{|p_j|\le \max{\{a,b\}}\atop |w|=z}|(\pa_p^wf)g(p_1,\ldots,p_n)|\,.
\ee
Then using the property of cutoff function in (\ref{cutoff}) 
one can find \cite{ckim3} that, for a fixed constant $\m$ of order $m$ ($\m$ 
may be considered as the momentum scale),
\bea
\norml{\pal\pa^z\lvl{r}{2n}}&&\no\\[3mm]
 &&\hspace{-25mm}\le \mbox{const}\Biggl\{(\LLL+m)
\norml{\pa^z\lvl{r}{2n+1}}\no\\
 &&\hspace{-22mm}+
   \sum_{r',l \atop z_1+z_2+z_3=z}
   \frac{1}{(\LLL+m)^{3+z_1}}\norml{\pa^{z_2}\lvl{r'}{2l}}
        \norml{\pa^{z_3}\lvl{r-r'}{2(n-l+1)}}\Biggr\}\,,\no\\
&&\label{fvlbh}
\eea
where ``const'' stands for some finite number which is independent of
$\LLL$ and $\lo$ (but may have dependence on $m$ and $\m$ through
$(\m/m)$). This estimate shows that the derivative of
vertex functions with respect to $\LLL$ is well bounded for all 
$\LLL\in[0,\lo]$.

As a next step, we apply this estimate to the integral form of the flow
equations. For irrelevant vertices, $2n+z>4$, boundary conditions are
given at $\LLL=\lo$ and so we integrate the flow equation from $\LLL=\lo$ down
to $\LLL$:
\bea \label{pf11}
\norml{\pa^z\lvl{r}{2n}}
 &=&\norml{\pa^z(\lvl{r}{2n}-\lvlo{r}{2n})}\no\\
 &\le&\int_\LLL^\lo d\lll \norm{\pa_\lll\pa^z\lvll{r}{2n}}{(2\lll,\m)}\,.
\eea
For the relevant vertices, i.e., in the case of $2n+z\le4$, 
we integrate the flow equation from 0 to $\LLL$ (at $p_i=0$) instead,
\be \label{pf111}
|\pa_p^w\lvl{r}{2n}(0)-\pa_p^w\lvo{r}{2n}(0)|
 \le\int_0^\LLL d\lll \norm{\pa_\lll\pa^z\lvll{r}{2n}}{(2\lll,\m)}\,.
\ee
The right hand side of (\ref{pf11}) and (\ref{pf111}) are now estimated 
and integrated using (\ref{fvlbh}). Finally, applying a suitable induction
argument over the perturbation order $r$ and the number of legs $n$, 
we obtain the following bound on vertex functions \cite{ckim3}
\be \label{th1a}
\norml{\pa^z\lvl{r}{2n}}\le(\LLL+m)^{4-2n-z}\plog{\LLL+m}{m}\,,\qquad
                        0\le\LLL\le\lo\,,
\ee
where {\rm Plog$((\LLL+m)/m)$} is some polynomial in $\log((\LLL+m)/m)$
whose coefficients are independent of $\LLL$ and $\lo$.
In particular, at $\LLL=0$, (\ref{th1a}) tells us that amputated 
connected Green functions and their derivatives of the theory are bounded by
\be
|\pa_p^wG_{r,2n}^{c}(p_1,\ldots,p_{2n-1})|
      \le \mbox{const}\cdot m^{4-2n-|w|}\,,\qquad \mbox{for }|p_i|\le\m\,.
\ee

Convergence of Green functions as $\lo\ra\infty$ can also be proved in a
similar way and the result is
\be\label{th2}
\norml{\pa_\lo\pa^z\lvl{r}{2n}}
     \le\left(\frac{\LLL+m}{\lo}\right)^3(\LLL+m)^{3-2n-z}\plog{\lo}{m}\,,
     \qquad 0\le\LLL\le\lo\,.
\ee
Consequently, amputated connected Green functions
$\pa_p^wG_{r,2n}^{c}$ converge to finite limits as fast as
$O[(\frac{m}{\lo})^2]$ (modulo powers of $\log(\lo/m)$). 

Before we move to the next topic we make a few remarks.

(i) When estimating the relevant vertices, we have had to 
integrate up from $\LLL=0$ and not just down
from $\lo$. If we had integrated down from $\lo$ assuming ``natural values''
like the irrelevant-vertex case, we would have had simply
$\norml{\lvl{r}{2}}\le\lo^2\plog{\lo}{\LLL+m}$ etc. That is, by imposing Eq.\
(\ref{effbc}) we have forced the initial bare values of $\r_a^0$ to be finely
adjusted so as to produce a scalar with mass $m\ll\lo$. This is the famous
naturalness problem in scalar theories \cite{naturalness}.

(ii) In this proof of perturbative renormalizability, we have not 
encountered any complication usually found in diagrammatic methods, for
example, overlapping divergences, cancellation between diagrams and so on. 
This is because, here, we always work with the whole Green function
directly and so problems in considering only a part of a Green
function do not appear.

\subsection{The Full Theory}

We are now going to discuss decoupling and heavy-mass factorization of
effective field theories using the renormalization group flow equation.
We will try to avoid going into technical details but explain only main line
of the argument. As our model for the full theory, We consider a scalar theory
($\p$-$\ps$ theory)
which involves two real scalar fields $\p$ and $\ps$ of which the masses are
$m$ and $M(\gg\lo)$, respectively, and interact via quartic couplings. Let us
write the bare action as
\be
S^{(f)\lo}=\frac12\lbra\p,\dmlo^{-1}\p\rket+\frac12\lbra\ps,\dMlo^{-1}\ps\rket
      +\lfilo\,.
\ee
Here, $\dmlo$ and $\dMlo$ are respective free-particle cutoff propagators,
and the interaction part $\lfilo$ is given by
\bea \label{sbare}
\lfilo&=&\intdx\left[\frac12\r_1^f\p^2(x)+\frac12\r_2^f(\pa_\m\p(x))^2
     +\frac1{4!}\r_3^f\p^4(x)+\frac12\r_4^f\ps^2(x)\right. \no\\
     & &\left.\hspace{15mm}+\frac12\r_5^f(\pa_\m\ps(x))^2
     +\frac1{4!}\r_6^f\ps^4(x)+\frac14\r_7^f\p^2\ps^2(x)\right]\,.
\eea
The superscript $f$ is used for couplings of the full theory.
As in the previous section, the bare couplings $\r_a^f$ ($a=1,2,\cdots,7)$
may be written as
\be \label{barefc}
\begin{array}{lll}
\r_1^f=Z_\p m_0^2-m^2,&\r_2^f=Z_\p-1,&\r_3^f=Z_\p^2 g_1^0,\\[2mm]
\r_4^f=Z_\ps M_0^2-M^2,&\r_5^f=Z_\ps-1,&\r_6^f=Z_\ps^2 g_2^0,\\[2mm]
\r_7^f=Z_\p Z_\ps g_3^0\,,&&
\end{array}
\ee
where $(m_0^2,M_0^2)$, $(Z_\p,Z_\ps)$ and $(g_1^0,g_2^0,g_3^0)$ are bare
masses, wave-function renormalizations and bare coupling constants,
respectively.

As pointed out in section 2, there is a freedom to choose the 
cutoff functions in propagators $\dml$ and $\dMl$. 
Here, we take advantage of this flexibility
and choose a ``mass-dependent scheme,'' i.e., take different
cutoff functions for the light-particle propagator $\dml$ and the
heavy-particle propagator $\dMl$: $\dml$ is chosen to be the same as Eq.\
(\ref{lightp}) while, for $\dMl$, we choose
\be \label{heavyp}
\dMlo=\frac{R_M(\lo,p)}{p^2+M^2}\,,
\ee
with
\be \label{heavycutoff}
R_M(\LLL,p)=\left[1-K\left(\LLL^2/M^2\right)\right]
               K\left(\frac{p^2}{\LLL^2}\right)\,.
\ee
The reason for this choice is because, if $\LLL<M$, we have $R_M=0$
identically. This property is very natural for our purpose, since it implies
that at the scale $\LLL=M$ we have all modes of the heavy field
$\ps$ integrated out and there remains only the light field $\p$ below
the scale $M$; it
explicitly implements Wilson's point of view on effective field theory.
Therefore, in the flow equation,
terms involving the heavy-particle propagator drop out if
$\LLL<M$ and it will look like the flow equation for the theory having the
light particle field only.
One may regard this property specific to our choice
(\ref{cutoff}) and (\ref{heavycutoff}) as the
analogy of the so-called ``manifest decoupling'' in
the conventional approach \cite{collins}. On the contrary, if one chose a
``mass-independent scheme'', i.e. if $R_m$ and $R_M$ were chosen independently
of their masses $m$ and $M$, a substantial part of heavy particle modes would
not be integrated out even below the scale $M$, thus making
subsequent discussions rather complicated.

Now, we define $\lfil[\p,\ps]$ following the general line discussed in
section 2. It will then satisfy the flow equation,
\bea \label{fflow}
&&\pal\lfil=-\frac12\intdp\left\{\pal\dml(p)
       \left[\frac{\d^2\lfil}{\d\p(p)\d\p(-p)}
      -\frac{\d\lfil}{\d\p(p)}\frac{\d\lfil}{\d\p(-p)}\right]\right.\no\\
&&\hspace{40mm}+\left.(m\ra M,\ \p\ra\ps)\Biggr\}
      \right|_{\rm field\mbox{-}dep.\ part}\,.
\eea
Expanding $\lfil$ in powers of $\p$ and $\ps$ in momentum space, we write
\bea\label{expansion}
\lfil[\p,\ps]=\sum_{|r|\ge0}\sum_{|n|\ge1}g^r
     \int&&\prod_{j=1}^{2n_1}\frac{d^4p_j}{(2\pi)^4}
      \prod_{j=1}^{2n_2-1}\frac{d^4p'_j}{(2\pi)^4}
      \p(p_1)\cdots\p(p_{2n_1})\ps(p'_1)\cdots\ps(p'_{2n_2-1})\no\\
    &&\times\lfl{r}{2n}(p_1,\ldots,p_{2n_1};p'_1,\ldots,p'_{2n_2-1})\,.\\
 &&\hspace{-15mm}{\textstyle \left(p'_{2n_2}\eq-\sum_{j=1}^{2n_1}
    p_j-\sum_{j=1}^{2n_2-1}p'_j\right)}\no
\eea
Some explanations on our notations are in order. $g\eq(g_1,g_2,g_3)$ are
perturbation-expansion parameters which may be identified as renormalized
coupling constants, $r$ and $n$ represent $(r_1,r_2,r_3)$
and $(n_1,n_2)$, respectively, with $|n|\eq n_1+n_2$, $|r|\eq r_1+r_2+r_3$
and finally, $g^r\eq g_1^{r_1}g_2^{r_2}g_3^{r_3}$.
The vertex functions satisfy the flow equation
\bea \label{lflflow}
\lefteqn{\pal\lfl{r}{2n}(p_1,\ldots,p_{2n_1};p'_1,\ldots,p'_{2n_2-1})}\no\\
&&=-\pmatrix{2n_1+2\cr 2}\intdq \pal\dml(q)
 \lfl{r}{2(n_1+1,n_2)}(p_1,\ldots,p_{2n_1},q,-q;p'_1,\ldots,p'_{2n_2-1})\no\\
&&\ +2\sum_{r'+r''=r\atop {l_1+l'_1=n_1+1\atop
l_2+l'_2=n_2}}l_1l'_1\pal\dml(P)
 \left[\lfl{r'}{2(l_1,l_2)}(p_1,\ldots,p_{2l_1-1},P;p'_1,\ldots,p'_{2l_2-1})
\right.\no\\[-7mm]
&&\left.\hspace{47mm}\times
       \lfl{r''}{2(l'_1,l'_2)}(p_{2l_1},\ldots,p_{2n_1},P;
                     p'_{2l_2+1},\ldots,p'_{2n_2-1})\right]_{symm} \no\\
&&\ -\pmatrix{2n_2+2\cr 2}\intdq \pal\dMl(q)
 \lfl{r}{2(n_1,n_2+1)}(p_1,\ldots,p_{2n_1};p'_1,\ldots,p'_{2n_2},q)\no\\
&&\
 +2\sum_{r'+r''=r\atop {l_1+l'_1=n_1\atop l_2+l'_2=n_2+1}}l_2l'_2\pal\dMl(P')
 \left[\lfl{r'}{2(l_1,l_2)}(p_1,\ldots,p_{2l};p'_1,\ldots,p'_{2l_2-1})
\right.\no\\[-7mm]
&&\left.\hspace{49mm}\times
       \lfl{r''}{2(l'_1,l'_2)}(p_{2l_1+1},\ldots,p_{2n_1};
                     p'_{2l_2},\ldots,p'_{2n_2})\right]_{symm}\,,
\eea
where $P=-\sum_{j=1}^{2l_1-1}p_j-\sum_{j=1}^{2l_2}p'_j$,
$P'=-\sum_{j=1}^{2l_1}p_j-\sum_{j=1}^{2l_2-1}p'_j$.
This equation can be represented by the
diagram shown in Fig.\ 2, where we denote $\p$($\ps$)-legs by
thin(thick)-lines. Also, 
With our choice of propagators,
(see Eq.\ (\ref{heavycutoff})), the last two terms in Eq.\ (\ref{lflflow})
vanish for
$\LLL<M$. Therefore the form of this flow equation coincides with that of a
field theory for a single scalar field.
Finally, in relation with the low-energy effective
theory, it should be noted that,
for $\lfl{r}{2(n_1,0)}$ (i.e., no external heavy particles),
the last term in Eq.\ (\ref{lflflow})
is identically zero and the heavy field $\ps$ enters
the flow equation only through the $\LLL$-differentiated propagator
in the third term.
\begin{figure}
\setlength{\unitlength}{1mm}
\begin{picture}(0,0)(-20,-60)
\put(22.5,0){\makebox(0,0)[cc]{\LARGE =}}
\put(70,0){\makebox(0,0)[cc]{\LARGE +}}
\put(7,-40){\makebox(0,0)[cc]{\LARGE +}}
\put(15,-40){\makebox(0,0)[cc]{$\displaystyle\sum_{l}$}}
\put(70,-40){\makebox(0,0)[cc]{\LARGE +}}
\put(78,-40){\makebox(0,0)[cc]{$\displaystyle\sum_{l}$}}

\thicklines
\put(0,0){\circle{10}}
\put(-4.5,-2.25){\line(-2,-1){5.06}}\put(-4.65,-2.10){\line(-2,-1){5.06}}
\put(-4.35,-2.40){\line(-2,-1){5.06}}
\put(4.5,-2.25){\line(2,-1){5.06}}\put(4.65,-2.10){\line(2,-1){5.06}}
\put(4.35,-2.40){\line(2,-1){5.06}}
\put(-3.74,-3.74){\line(-1,-1){4}}\put(-3.89,-3.61){\line(-1,-1){4}}
\put(-3.60,-3.90){\line(-1,-1){4}}
\put(-2.5,-7){\makebox(0,0)[cc]{$\cdot$}}
\put(0,-7.43){\makebox(0,0)[cc]{$\cdot$}}
\put(2.5,-7){\makebox(0,0)[cc]{$\cdot$}}
\put(4.7,-5.8){\makebox(0,0)[cc]{$\cdot$}}
\put(0,-12){\makebox(0,0)[cc]{$\sf 2n_2$}}
\put(-15,0){\makebox(0,0)[cc]{\LARGE$\partial_\Lambda$}}
\thinlines
\put(4.5,2.25){\line(2,1){5.06}}
\put(-4.5,2.25){\line(-2,1){5.06}}
\put(-3.7,3.7){\line(-1,1){4}}
\put(-2.5,7){\makebox(0,0)[cc]{$\cdot$}}
\put(0,7.43){\makebox(0,0)[cc]{$\cdot$}}
\put(2.5,7){\makebox(0,0)[cc]{$\cdot$}}
\put(4.7,5.8){\makebox(0,0)[cc]{$\cdot$}}
\put(0,12){\makebox(0,0)[cc]{$\sf 2n_1$}}
\end{picture}
\begin{picture}(0,0)(-65,-60)
\thicklines
\put(0,0){\circle{10}}
\put(-4.5,-2.25){\line(-2,-1){5.06}}\put(-4.65,-2.10){\line(-2,-1){5.06}}
\put(-4.35,-2.40){\line(-2,-1){5.06}}
\put(4.5,-2.25){\line(2,-1){5.06}}\put(4.65,-2.10){\line(2,-1){5.06}}
\put(4.35,-2.40){\line(2,-1){5.06}}
\put(-3.74,-3.74){\line(-1,-1){4}}\put(-3.89,-3.61){\line(-1,-1){4}}
\put(-3.60,-3.90){\line(-1,-1){4}}
\put(-2.5,-7){\makebox(0,0)[cc]{$\cdot$}}
\put(0,-7.43){\makebox(0,0)[cc]{$\cdot$}}
\put(2.5,-7){\makebox(0,0)[cc]{$\cdot$}}
\put(4.7,-5.8){\makebox(0,0)[cc]{$\cdot$}}
\put(0,-12){\makebox(0,0)[cc]{$\sf 2n_2$}}
\thinlines
\put(4.5,2.25){\line(2,1){5.06}}
\put(-3.7,3.7){\line(-1,1){4}}
\put(-2.5,7){\makebox(0,0)[cc]{$\cdot$}}
\put(0,7.43){\makebox(0,0)[cc]{$\cdot$}}
\put(2.5,7){\makebox(0,0)[cc]{$\cdot$}}
\put(4.7,5.8){\makebox(0,0)[cc]{$\cdot$}}
\put(0,12){\makebox(0,0)[cc]{$\sf 2(n_1+1)$}}
\put(-13,7){\makebox(0,0)[cc]{$\sf\partial_\Lambda P_m$}}
\bezier{72}(-4.7,1.9)(-8.5,6.5)(-11,3)
\bezier{72}(-5,0)(-11.7,-1.7)(-11,3)
\end{picture}
\begin{picture}(0,0)(-110,-60)
\thicklines
\put(0,0){\circle{10}}
\put(4.5,-2.25){\line(2,-1){5.06}}\put(4.65,-2.10){\line(2,-1){5.06}}
\put(4.35,-2.40){\line(2,-1){5.06}}
\put(-3.74,-3.74){\line(-1,-1){4}}\put(-3.89,-3.61){\line(-1,-1){4}}
\put(-3.60,-3.90){\line(-1,-1){4}}
\put(-2.5,-7){\makebox(0,0)[cc]{$\cdot$}}
\put(0,-7.43){\makebox(0,0)[cc]{$\cdot$}}
\put(2.5,-7){\makebox(0,0)[cc]{$\cdot$}}
\put(4.7,-5.8){\makebox(0,0)[cc]{$\cdot$}}
\put(0,-12){\makebox(0,0)[cc]{$\sf 2(n_2+1)$}}
\put(-13,-6){\makebox(0,0)[cc]{$\sf\partial_\Lambda P_M$}}
\bezier{72}(-4.7,-1.9)(-8.5,-6.5)(-11,-3)\bezier{72}(-4.62,-2.2)(-8.3,-6.7)(-11.3,-3)
\bezier{72}(-5.05,0)(-11.7,1.7)(-11,-3)\bezier{72}(-5.05,0.19)(-11.9,1.9)(-11.3,-3)
\thinlines
\put(4.5,2.25){\line(2,1){5.06}}
\put(-4.5,2.25){\line(-2,1){5.06}}
\put(-3.7,3.7){\line(-1,1){4}}
\put(-2.5,7){\makebox(0,0)[cc]{$\cdot$}}
\put(0,7.43){\makebox(0,0)[cc]{$\cdot$}}
\put(2.5,7){\makebox(0,0)[cc]{$\cdot$}}
\put(4.7,5.8){\makebox(0,0)[cc]{$\cdot$}}
\put(0,12){\makebox(0,0)[cc]{$\sf 2n_1$}}
\end{picture}
\begin{picture}(0,0)(-45,-20)
\thicklines
\put(0,0){\circle{10}}
\put(-4.5,-2.25){\line(-2,-1){5.06}}\put(-4.65,-2.10){\line(-2,-1){5.06}}
\put(-4.35,-2.40){\line(-2,-1){5.06}}
\put(4.5,-2.25){\line(2,-1){5.06}}\put(4.65,-2.10){\line(2,-1){5.06}}
\put(4.35,-2.40){\line(2,-1){5.06}}
\put(-3.74,-3.74){\line(-1,-1){4}}\put(-3.89,-3.61){\line(-1,-1){4}}
\put(-3.60,-3.90){\line(-1,-1){4}}
\put(-2.5,-7){\makebox(0,0)[cc]{$\cdot$}}
\put(0,-7.43){\makebox(0,0)[cc]{$\cdot$}}
\put(2.5,-7){\makebox(0,0)[cc]{$\cdot$}}
\put(4.7,-5.8){\makebox(0,0)[cc]{$\cdot$}}
\put(0,-12){\makebox(0,0)[cc]{$\sf 2l_2$}}
\thinlines
\put(4.5,2.25){\line(2,1){5.06}}
\put(-4.5,2.25){\line(-2,1){5.06}}
\put(-3.7,3.7){\line(-1,1){4}}
\put(-2.5,7){\makebox(0,0)[cc]{$\cdot$}}
\put(0,7.43){\makebox(0,0)[cc]{$\cdot$}}
\put(2.5,7){\makebox(0,0)[cc]{$\cdot$}}
\put(4.7,5.8){\makebox(0,0)[cc]{$\cdot$}}
\put(0,12){\makebox(0,0)[cc]{$\sf 2l_1$}}
\put(5,0){\line(1,0){16.5}}
\put(13.5,3){\makebox(0,0)[cc]{$\sf\partial_\Lambda P_m$}}
\end{picture}
\begin{picture}(0,0)(-70,-20)
\thicklines
\put(0,0){\circle{10}}
\put(-4.5,-2.25){\line(-2,-1){5.06}}\put(-4.65,-2.10){\line(-2,-1){5.06}}
\put(-4.35,-2.40){\line(-2,-1){5.06}}
\put(4.5,-2.25){\line(2,-1){5.06}}\put(4.65,-2.10){\line(2,-1){5.06}}
\put(4.35,-2.40){\line(2,-1){5.06}}
\put(-3.74,-3.74){\line(-1,-1){4}}\put(-3.89,-3.61){\line(-1,-1){4}}
\put(-3.60,-3.90){\line(-1,-1){4}}
\put(-2.5,-7){\makebox(0,0)[cc]{$\cdot$}}
\put(0,-7.43){\makebox(0,0)[cc]{$\cdot$}}
\put(2.5,-7){\makebox(0,0)[cc]{$\cdot$}}
\put(4.7,-5.8){\makebox(0,0)[cc]{$\cdot$}}
\put(0,-12){\makebox(0,0)[cc]{$\sf 2(n_2-l_2)$}}
\thinlines
\put(4.5,2.25){\line(2,1){5.06}}
\put(-4.5,2.25){\line(-2,1){5.06}}
\put(-3.7,3.7){\line(-1,1){4}}
\put(-2.5,7){\makebox(0,0)[cc]{$\cdot$}}
\put(0,7.43){\makebox(0,0)[cc]{$\cdot$}}
\put(2.5,7){\makebox(0,0)[cc]{$\cdot$}}
\put(4.7,5.8){\makebox(0,0)[cc]{$\cdot$}}
\put(0,12){\makebox(0,0)[cc]{$\sf 2(n_1-l_1+1)$}}
\end{picture}
\begin{picture}(0,0)(-105,-20)
\thicklines
\put(0,0){\circle{10}}
\put(-4.5,-2.25){\line(-2,-1){5.06}}\put(-4.65,-2.10){\line(-2,-1){5.06}}
\put(-4.35,-2.40){\line(-2,-1){5.06}}
\put(4.5,-2.25){\line(2,-1){5.06}}\put(4.65,-2.10){\line(2,-1){5.06}}
\put(4.35,-2.40){\line(2,-1){5.06}}
\put(-3.74,-3.74){\line(-1,-1){4}}\put(-3.89,-3.61){\line(-1,-1){4}}
\put(-3.60,-3.90){\line(-1,-1){4}}
\put(-2.5,-7){\makebox(0,0)[cc]{$\cdot$}}
\put(0,-7.43){\makebox(0,0)[cc]{$\cdot$}}
\put(2.5,-7){\makebox(0,0)[cc]{$\cdot$}}
\put(4.7,-5.8){\makebox(0,0)[cc]{$\cdot$}}
\put(0,-12){\makebox(0,0)[cc]{$\sf 2l_2$}}
\put(13.5,3){\makebox(0,0)[cc]{$\sf\partial_\Lambda P_M$}}
\put(5,-0.2){\line(1,0){16.5}}\put(5,0.2){\line(1,0){16.5}}\put(5,0){\line(1,0){16.5}}
\thinlines
\put(4.5,2.25){\line(2,1){5.06}}
\put(-4.5,2.25){\line(-2,1){5.06}}
\put(-3.7,3.7){\line(-1,1){4}}
\put(-2.5,7){\makebox(0,0)[cc]{$\cdot$}}
\put(0,7.43){\makebox(0,0)[cc]{$\cdot$}}
\put(2.5,7){\makebox(0,0)[cc]{$\cdot$}}
\put(4.7,5.8){\makebox(0,0)[cc]{$\cdot$}}
\put(0,12){\makebox(0,0)[cc]{$\sf 2l_1$}}
\end{picture}
\begin{picture}(140,74)(-130,-20)
\thicklines
\put(0,0){\circle{10}}
\put(-4.5,-2.25){\line(-2,-1){5.06}}\put(-4.65,-2.10){\line(-2,-1){5.06}}
\put(-4.35,-2.40){\line(-2,-1){5.06}}
\put(4.5,-2.25){\line(2,-1){5.06}}\put(4.65,-2.10){\line(2,-1){5.06}}
\put(4.35,-2.40){\line(2,-1){5.06}}
\put(-3.74,-3.74){\line(-1,-1){4}}\put(-3.89,-3.61){\line(-1,-1){4}}
\put(-3.60,-3.90){\line(-1,-1){4}}
\put(-2.5,-7){\makebox(0,0)[cc]{$\cdot$}}
\put(0,-7.43){\makebox(0,0)[cc]{$\cdot$}}
\put(2.5,-7){\makebox(0,0)[cc]{$\cdot$}}
\put(4.7,-5.8){\makebox(0,0)[cc]{$\cdot$}}
\put(0,-12){\makebox(0,0)[cc]{$\sf 2(n_2-l_2+1)$}}
\thinlines
\put(4.5,2.25){\line(2,1){5.06}}
\put(-4.5,2.25){\line(-2,1){5.06}}
\put(-3.7,3.7){\line(-1,1){4}}
\put(-2.5,7){\makebox(0,0)[cc]{$\cdot$}}
\put(0,7.43){\makebox(0,0)[cc]{$\cdot$}}
\put(2.5,7){\makebox(0,0)[cc]{$\cdot$}}
\put(4.7,5.8){\makebox(0,0)[cc]{$\cdot$}}
\put(0,12){\makebox(0,0)[cc]{$\sf 2(n_1-l_1)$}}
\end{picture}
\caption{Graphical representation of the flow equation (\ref{lflflow}).
Thin (thick) lines represents light (heavy) field.}
\end{figure}
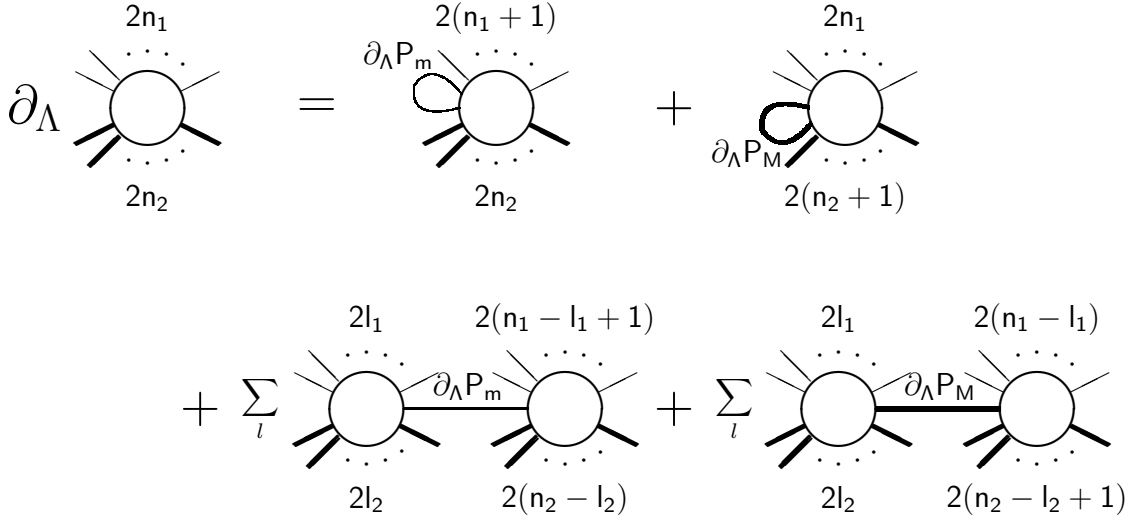

The boundary conditions which $\lfl{r}{2n}$'s obey are given as in the
previous section. First, at $\LLL=\lo$, irrelevant vertices vanish, i.e.,
\be \label{fbc}
\pa_p^w\lflo{r}{2n}=0,\hspace{45mm}2|n|+|w|>4\,.
\ee
At $\LLL=0$, we impose renormalization conditions on relevant terms:
\be \label{fbc2}
\begin{array}{lll}
\lfo{r}{(2,0)}(0)=0,&
 \pa_{p_\m}\pa_{p_\n}\lfo{r}{(2,0)}(0)=0,&\hspace{3mm}
 \lfo{r}{(4,0)}(0)=\frac1{4!}\d_{r,(1,0,0)}\,,\\[3mm]
\lfo{r}{(0,2)}(\bar p'_1)=0,
 & \pa_{p_\m}\pa_{p_\n}\lfo{r}{(0,2)}(\bar p'_1)=0,&\hspace{3mm}
 \lfo{r}{(0,4)}(\bar p'_2,\bar p'_3,\bar p'_4)
                        =\frac1{4!}\d_{r,(0,1,0)}\,,\\[3mm]
\lfo{r}{(2,2)}(0,0;\bar p'_5)=\frac14\d_{r,(0,0,1)}. &&
\end{array}
\ee
Here, normalization momenta ${\bar p'}_i$ ($i=1,\ldots,5$) are chosen to  be
constants of magnitude $M$; i.e.,
we have chosen the renormalization points for the light-particle Green
functions at zero momentum, and those for the heavy particles at momenta of
magnitude $M$.

With the flow equation (\ref{lflflow}) and the boundary conditions (\ref{fbc})
and (\ref{fbc2}), now one can demonstrate the perturbative renormalizability
(i.e., boundedness and convergence of Green functions as $\lo\ra\infty$) of
the full theory following the line of arguments given in section 3.1, with 
the help of our choice of cutoff functions. The result is \cite{ckim3,ball2}:
for vertices with no external heavy-particle leg,
\be \label{11a}
\normn{\pa^z\lfl{r}{2(n_1,0)}}
     \le(\LLL+m)^{4-2n_1-z}\plog{\LLL+m}{m}\,,
\ee
while, for irrelevant vertices ($2|n|+z>4$) with external heavy-particle legs,
\be \label{11b}
\normn{\pa^z\lfl{r}{2n}}
     \le(\LLL+m)^{4-2|n|-z}\plog{\LLL+M}{m}\,,
\ee
and, finally, for relevant vertices ($2|n|+4\le4$) with external
heavy-particle legs,
\be \label{11c}
\normn{\pa^z\lfl{r}{2n}}
     \le(\LLL+M)^{4-2|n|-z}\plog{\LLL+M}{m}\,,
\ee
where the coefficients in polynomials of logarithms are independent of $M$,
$\LLL$, and $\lo$, and the norm $\normn{\cdot}$ is defined by
\be \label{newnorm}
\norm{\pa^z\lfl{r}{2(n_1,n_2)}}{(a,b,b')}
  \eq\max_{|p_i|\le\max\{a,b\}\atop{|p'_i|\le\max\{a,b'\}\atop z=|w|}}
 |\pa_p^w\lfl{r}{2(n_1,n_2)}(p_1,\ldots,p_{2n_1};p'_1,\ldots,p'_{2n_2-1})|\,.
\ee
(If there is no
external heavy-particle leg ($n_2=0$), this definition reduces to that of the
norm $\|\ \|_{(a,b)}$ in Eq.\ (\ref{norm}).)

Actually these bounds are not unexpected --- they just show the
right scaling behaviors. That is, the effect of imposing
the renormalization conditions at momenta of magnitude $M$ for heavy-particle
legs shows with appropriate powers of $M$ (up to logarithmic corrections)
in the flow equation of vertices with external heavy-particle legs.
On the other hand, in the light-particle sector, the
bounds have the same form as those of the single scalar theory (see Eq.\
(\ref{th1a})): the large $M$ corrections do not show up in the bounds because
the light-particle mass $m$ is forced to be (unnaturally) small by hand.
Also, it should be noted that the bounds (\ref{11a}) on the light
particle sector have exactly the same form as that of the theory with a single
scalar field. Of course, this should be the case for decoupling to
occur in the first place.

\subsection{Low-Energy Effective Theory}
Now suppose that we are interested in physics at scale much smaller that $M$.
What is the low-energy effective theory? Decoupling theorem states that, to the
zeroth order of $1/M$, it is simply given by the $\p^4$-theory with heavy
field $\psi$ removed from the Lagrangian of the full theory. To establish
this, we should show that the difference between Green functions of the
$\p$-$\ps$ theory
and those of the $\p$ theory are at most of order $1/M^2$ (no $O(1/M)$ 
term because of the $Z_2$ symmetry). In our approach,
it is given by the following bound:
\bea\label{th3}
\lefteqn{\norml{\pa^z(\lfl{r}{2(n,0)}-\lvl{r_1}{2n}\d_{r,(r_1,0,0)})}}\no\\
&&\le\left\{\begin{array}{ll}
        \displaystyle\left(\frac{\LLL+m}{M}\right)^2
                      (\LLL+m)^{4-2n-z}\plog{M}{m}\,,&
                      \hspace{2mm} 0\le\LLL\le M\\[4mm]
        \displaystyle\LLL^{4-2n-z}\plog{\LLL}{m}\,,&
                      M\le\LLL\le\lo\,,\end{array}\right.
\eea
where $r=(r_1,r_2,r_3)$.
Notice that the difference in the vertex functions is no longer small if $\LLL$
becomes comparable to the heavy particle mass $M$, which implies that
low-energy effective theory is not useful above the heavy particle mass
scale. 

The strategy to show this bound is almost straightforward. As we have seen,
the form of the flow equation of $\p$-$\ps$ theory is the same as that of 
$\p$ theory for $\LLL<M$. Moreover, the boundary conditions of the two
theories 
are the same by definition. Consequently, considering the flow of the
difference of vertex functions,
\be
\dvl{r}{2n}\eq\lfl{r}{2(n,0)}-\lvl{r_1}{2n}\d_{r,(r_1,0,0)}\,.
\ee
we can expect that $\dvl{r}{2n}$ is almost zero since boundary
conditions for $\dvl{r}{2n}$ is zero and the flow equation for $\dvl{r}{2n}$
is homogeneous for $\LLL<M$. Following the steps explained in the previous
section, one can easily establish the bound (\ref{th3}). As we expect,
decoupling theorem is proved more or less trivially in this approach.
\vspace{5mm}

The zeroth order effective theory is just renormalizable $\p$ theory.
In other words, it describes the light particle physics accurately to the
zeroth order of $1/M$ at low energy.
Since it is perturbatively renormalizable, the theory itself does 
not know the scale $M$ below which the theory is effective.
In a sense the effective theory nature becomes manifest only when we
raise the accuracy to higher order in $1/M$. It is then necessary to include
irrelevant (nonrenormalizable) operators in the effective Lagrangian. In
addition, we must also supply appropriate calculation rules
to obtain unambiguous finite results because the presence of such irrelevant
terms can make physical quantities
diverge if naively calculated, as we mentioned before.
If these tasks are systematically performed with
local operators, we will have a local effective quantum field theory
where virtual heavy-particle effects are isolated into the coefficients of
irrelevant operators in the Lagrangian to any desired order of accuracy.
Let us suppose that we want to describe low-energy physics
in the full theory
accurately up to order $1/M^{2N_0}$. ($N_0$ is some positive integer.)
We will show below that we can then factorize all heavy-particle effects to
the given order theory by making use of the flow equation (\ref{flowk}) 
and by appropriately choosing the operators $\lilon$, $N=1,2,\ldots,N_0$.

Here, $\lilon$'s may be assumed to have the form
\be \label{lilon}
\lilon=\intdx
      \pmatrix{\mbox{local, Euclidean invariant, even polynomials}\cr
      \mbox{in $\p$ and its derivatives, of dim.$\le 4+2N$}}
\,,\hspace{5mm} N=1,2,\ldots,N_0\,.
\ee
The coefficient of the polynomials will be chosen later so that $\lilon$
carries information appropriate to $O(1/M^{2N})$-effects from the full theory.
Then we claim that the sum of $\lilon$'s, viz.,
\be \label{ltilon}
\ltilog{N_0}\eq\lilo+\sum_{N=1}^{N_0}\frac{\lilon}{N!}
   \ \Bigl(=\ltilog{N_0-1}+\frac{\lilog{N_0}}{N_0!}\Bigr)\,,
\ee
will reproduce the predictions of the full theory at low energy with an
accuracy of order $1/M^{2N_0}$. As we noted in section 2, 
$\lilon$ will describe
amputated connected Green functions with the insertion of the operator
$O_{N}$ defined in Eq.\ (\ref{oiexpansion}).
As usual, the quantity  $\liln$, which satisfies the flow equation
(\ref{flowk}), may be expanded in powers of $\p$:
\bea \label{lilnexpansion}
&&\liln[\p]=\sum_{r=0}^\infty\sum_{n=1}^\infty g_1^r
         \int\prod_{j=1}^{2n-1}\frac{d^4p_j}{(2\pi)^4}\p(p_1)\cdots \p(p_{2n})
       \lvln{r}{2n}(p_1,\ldots,p_{2n-1})\,,\\
&&\hspace{80mm}
  {\textstyle\left(p_{2n}\eq-\sum_{j=1}^{2n-1}p_j\right)\,.}\no
\eea
We may also define $\ltln{r}{2n}$'s through the same kind of relation as in
Eq.\ (\ref{ltilon}). To complete the definition of $\lilon$, we must also
state the boundary
conditions for $\lvln{r}{2n}$. At $\LLL=\lo$, Eq.~(\ref{lilon}) implies that
\be \label{lvlnbc1}
 \pa_p^w\lvlon{r}{2n}=0,\quad \mbox{for }2n+|w|>4+2N\,.
\ee
For $2n+|w|\le4+2N$, we impose the following condition, recursively in $N$:
\be \label{lvlnbc2}
 \pa_p^w\lvon{r}{2n}(0)=N!\,\pa_p^w(\lbo{r}{2n}(0)-\ltog{N-1}{r}{2n}(0))\,,
\ee
where $\bar{\cal G}$ is the vertex function obtained by summing over
the couplings involving heavy fields in Eq.~(\ref{expansion}), i.e.,
\be\label{lbl}
\lbl{r_1}{2n}\eq\sum_{r_2,r_3}g_2^{r_2}g_3^{r_3}\lfl{(r_1,r_2,r_3)}{2n}\,.
\ee
As a consequence, $\pa_p^w\lvon{r}{2n}(0)=0$ for the case $2n+|w|<4+2N$.
The meaning of this boundary condition is clear. To obtain the $N$-th order
effective theory, we add new (higher dimensional) operators to the effective 
action which compensate the difference between the full theory and the 
$N-1$-th order effective theory. 

As we discussed in section 1, this seemingly natural procedure has a 
potential problem because new operators
are nonrenormalizable and naive treatment of those operators make Green 
functions divergent (rather than making the effective theory more accurate).
However one can show that, if one demands the flow of the effective 
theory follow Eq.\ (\ref{flowk}), 
divergences are automatically taken care of and virtual
heavy particle effects are correctly incorporated. Indeed, straightforward
extension of the argument used in proving decoupling shows that \cite{ckim3}
the difference between two  vertex functions, $\lfl{r}{2(n,0)}$ of the
$\p$-$\ps$ theory and $\ltpln{r}{2n}$ of the $\p$ theory, satisfies the bound
\be\label{th4}
\norml{\pa^z(\lfl{r}{2(n,0)}-\ltpln{r}{2n})}\hspace{-2mm}
\le\left\{\begin{array}{ll}
          \displaystyle\left(\frac{\LLL+m}{M}\right)^{2N+2}\hspace{-3mm}
                       (\LLL+m)^{4-2n-z}\plog{M}{m}\,,&
                        0\le\LLL\le M\\[4mm]
          \displaystyle\left(\frac{\LLL}{M}\right)^{2N}
                        \LLL^{4-2n-z}\plog{\LLL}{m}\,,&
                        M\le\LLL\le\lo\,,\end{array}\right.
\ee
which establishes the factorization of virtual heavy particle effects to any
given order of accuracy in $1/M$,
\bea
&&|\pa_p^w(G_{r,2(n,0)}^{(f)c}(p_1,\ldots,p_{2n-1})
             -\widetilde G_{r,2n}^{'c;N_0}(p_1,\ldots,p_{2n-1}))|\no\\
 &&\hspace{43mm}\le\left(\frac{m}{M}\right)^{2N_0+2}m^{4-2n-z}\plog{M}{m},\no\\
&&\hspace{80mm}(\mbox{for } |p_i|\le\m)\,.
\eea
If $\LLL$ is larger than $M$, Eq.\ (\ref{th4}) says
that the vertex function $\pa_p^w\ltln{r}{2n}$ is larger than
$\pa_p^w\ltlg{N=0}{r}{2n}$ by a factor $(\LLL/M)^{2N}$. Therefore, in this
scale, increasing $N$ (or adding more irrelevant terms, equivalently) makes
the vertex functions {\it unnaturally} large and the theory becomes ``less
effective''. This is because we have imposed
unnatural renormalization conditions
(\ref{lvlnbc2}) on irrelevant vertices. If we chose natural values as the bare
irrelevant couplings, they would give only $O((m/\lo)^{2N})$ contributions to
Green functions; but, in Eq.\ (\ref{lvlnbc2}), we imposed the
values of order $(m/M)^{2N}$ to $\ltln{r}{2n}$ as the renormalization
conditions, which are natural only if the cutoff is around $M$. This affirms
that the effective theory is useful only below the heavy particle mass
scale $M$.
\vspace{5mm}

So far we have shown that, at low energy, connected amputated
Green functions of the full theory are reproduced by considering those of
the $\p$ theory plus those with insertion of operators $O_{N}$,
$N=1,\cdots,N_0$, defined in Eq.\ (\ref{oiexpansion}).
Such operator insertions have a simple interpretation if one
uses the notion of renormalization of composite operators and their normal 
product notation. As mentioned in section 2 this subject can also be easily
described using the flow equation but we will not discuss it in this lecture.
Interested reader can consult Refs. \cite{keller,ckim3}. 
Results are the following. For $N_0=1$, $\lilog{1}$ can be written as
\be \label{lilo1}
\lilog{1}=\sum_{2n+|w|=6}b_{2n,\{w\}}^{R;1}[\mono{2n}{w}]\,,
\ee
where
$b_{2n,\{w\}}^{R;1}
 =\frac{1}{\caln}\pa_p^w(G_{2(n,0)}^{(f)c}-G_{2n}^c)(0)$, $\caln$ is a
combinatorial factor defined by 
\be \label{caln}
\left.\pa_p^{w'}\left[(ip_1)^{w_1}\cdots(ip_{2n})^{w_{2n}}\right]_{symm}
        \right|_{p_1=\cdots=p_{2n}=0}=\d_{\{w\}\{w'\}}\caln\,,
\ee
and $[\mono{2n}{w}]$ is the normal product of a local vertex
$\mono{2n}{w}=\intdx \pa_x^{w_1}\p\cdots\pa_x^{w_{2n-1}}\p\p(x).$
In terms of the local
Euclidean-invariant vertices of dimension six,
the above equation can be cast into a more attractive form:
\bea \label{lilo11}
\lilog{1}
 &=&a_1^{(6)}\intdx[\p(\pa^2)^2\p(x)]+a_2^{(6)}\intdx[\p^3\pa^2\p(x)]
        +a_3^{(6)}\intdx[\p^6(x)]\no\\
 &\eq&\sum_{i=1}^3a_i^{(6)}[\O_i^{(6)}]\,,
\eea
with
\bea \label{ais}
a_1^{(6)}&=&\frac{1}{8\cdot4!}(\pa^2)^2(G_{(2,0)}^{(f)c}-G_2^c)(0)\,,\no\\
a_2^{(6)}&=&-\frac{1}{8}\pa^2(G_{(4,0)}^{(f)c}-G_4^c)(0)\,,\no\\
a_3^{(6)}&=&(G_{(6,0)}^{(f)c}-G_6^c)(0)\,.
\eea
The $1/M^2$-order effects of the full theory are then described by the
insertion of the operator $O_1=\sum_i a_i^{(6)}[\O_i^{(6)}]$, viz.,
\be
G_{2(n,0)}^{(f)c}=G_{2n}^c+\sum_{i=1}^3a_i^{(6)}G_{2n}^{c;1}([\O_i^{(6)}])
  +O\left(\frac{m^{8-2n}}{M^4}\plog{M}{m}\right)\,,
\ee
where $G_{2n}^{c;1}([\O_i^{(6)}])$ denotes the Green function with
$[\O_i^{(6)}]$ inserted.
For general $N_0$, $\lilon$ $(N=1,\ldots,N_0$) can be identified as
\be \label{lilon1}
\lilon=\sum_{2n+|w|=4+2N}b_{2n,\{w\}}^{R;N}[\mono{2n}{w}]+\lilonf_{CT}\,,
\ee
where
\be
b_{2n,\{w\}}^{R;N}
 =\frac{N!}{\caln}\sum_{2n+|w|=4+2N}
  \pa_p^w(G_{2(n,0)}^{(f)c}-\widetilde G_{2n}^{c;N-1})(0)\,,
\ee
and $\lilonf_{CT}$ denote counterterms which are needed to cancel the new
divergences due to the multiple insertion of $\lilog{I}$'s,
$I=1,\ldots,N-1$.
(See \cite{ckim3} for explicit forms of $\lilonf_{CT}$.) 
Now let $[\O_i^{(4+2N)}]$'s form a  complete set of mutually independent
Euclidean-invariant local
vertices of dimension $(4+2N)$ and let $a_i^{(4+2N)}$'s be appropriate
expansion coefficients as in Eq.\ (\ref{lilo11}), so that we may write
\be \label{lilon2}
\lilon=\sum_ia_i^{(4+2N)}[\O_i^{(4+2N)}]+\lilonf_{CT}\,.
\ee
Then the effective action $\ltilog{N_0}$ may be written as
\bea
\ltilog{N_0}&=&\lilo+\sum_{N=1}^{N_0}
               \sum_i\frac{a_i^{(4+2N)}}{N!}[\O_i^{(4+2N)}]
               +\sum_{N=2}^{N_0}\frac{1}{N!}\lilonf_{CT}\no\\
    &=&\ltilog{N_0-1}+
        \sum_i\frac{a_i^{(4+2N_0)}}{N_0!}[\O_i^{(4+2N_0)}]
        +\frac{1}{N_0!}S^{\lo;N_0}_{CT}\,,
\eea
where $a_i^{(4+2N)}$'s are appropriate linear combinations of
$b_{2n,\{w\}}^{R;N}$'s (all of which are $O(1/M^{2N})$). Thus the
$1/M^{2N}$-order information of the full theory is factorized into the
coefficient $a_i^{(4+2N)}$'s of local vertices of dimension $(4+2N)$.
If we want to increase the
accuracy from $O(1/M^{2N_0})$ to $O(1/M^{2(N_0+1)})$, we have only to include
in $\ltilog{N_0}$ local vertices $[\O_i^{(4+2(N_0+1))}]$ of dimension
$4+2(N_0+1)$. (The lower dimensional vertices need not be modified.) The
coefficients $a_i^{4+2(N_0+1)}$ are calculable as the difference of
appropriate Green functions of the full theory and those of the effective
theory (i.e., $\ltilog{N_0}$) at renormalization point.
New divergences are subtracted away by
$S^{\lo;N_0+1}_{CT}$. Given this bare Lagrangian, the flow equation
guarantees that all Green functions are finite and accurate up to order
$1/M^{2(N_0+1)}$.

The interpretation of this result is clear from the viewpoint of the
renormalization group flow in the infinite dimensional space of possible
Lagrangians. As we reduce the cutoff, the bare Lagrangian of the full theory,
which is specified by the boundary conditions (\ref{fbc}) and (\ref{fbc2}),
flows down to a submanifold parametrized by relevant couplings only.
It has both heavy and light operators. Now the flow may be projected onto the
subspace of operators consisting
of the light field only. Then finding a local low-energy effective field
theory
of light particles is equivalent to finding a renormalization group flow with
a few (relevant or irrelevant) local operators which can
best approximate the projected flow of the full theory. To the zeroth order
in $1/M^2$, it is done with relevant terms only by choosing the same
renormalization conditions as those of the full theory. The deviation from the
full-theory flow is corrected by reading off values of the remaining
irrelevant
coordinates at the $\LLL=0$ point. At first, components of dimension six
operators may be read, which tell us the effects of order $1/M^2$;
the flow of
the full theory is now approximated to the order $1/M^2$ at $\LLL<M$. If one
wants to increase the accuracy, it is necessary to read more and more
irrelevant coordinates (at $\LLL=0$) of the projected flow of the full
theory and modify the effective-theory flow with the corresponding
flow equation given in
Eq.\ (\ref{flowk}). Our equation (\ref{flowk}) automatically takes care of
possible divergences due to the unnaturally large irrelevant components
in such a way that the bare Lagrangian
may contain only a finite number of irrelevant terms.

\section{Conclusions}

In the context of a simple scalar field theory, we have demonstrated that,
at low energy,
virtual heavy particle effects on light-particle Green functions are
completely
factorized via effective local vertices to any desired order. For this, we
have used the powerful flow equation approach which is
a differential version of Wilson's renormalization group transformation. 
In applying this method, we have seen that irrelevant terms in the Wilsonian
effective Lagrangian (which has a finite UV cutoff $\LLL=M$, a characteristic
scale representing heavy-particle thresholds) are replaceable by the
corresponding higher
dimensional composite operators plus counterterms for their products in the
continuum limit (where UV cutoff is supposed to go to infinity).
The latter can be dealt with with the help of the normal product concept.
Once this fact is noticed, factorization is straightforward:
all $1/M^{2N}$-order effects can be isolated in terms of local vertices
involving operators of dimension $(4+2N)$. Thereby we arrive at
a local continuum effective field theory which describes low-energy 
light-particle
physics accurately up to any desired order in $1/M^2$ with appropriate
calculation rules for irrelevant (nonrenormalizable) vertices.
Since the arguments here are essentially
dimensional and largely theory-independent, it should not be difficult
to generalize them to different field theories such as
gauge theories or theories with spontaneous symmetry breaking.
Indeed, decoupling was proved
for these theories in \cite{girardello} within this approach.

Since the couplings for effective local vertices contain powers of $\log
(M/m)$
in perturbation theory, it is often desirable to sum such logarithms in a
systematic way. This has been achieved by utilizing a set of improved
Callan-Symanzik equations \cite{clee}. Also, here we limited our attention to
vertex functions at momentum range $\m=O(m)$.
As one increases $\m$ to sufficiently high energy scale,
the $(\m/m)$-dependence we neglected will become important.
Keeping such $\m$-dependence and establishing bounds more carefully, one
may study the high-momentum behavior of Green functions (even in the
exceptional momentum region) within the effective field theory context.
This bound has been obtained in \cite{ball2} for zeroth order
effective theory (corresponding to decoupling). 

In summary, the flow equation, which is based on Wilsonian viewpoint of 
effective theory, provides clear understanding of effective theories in
continuum.

\acknowledgments
It is my great pleasure to thank Prof. D.\ P.\ Min for inviting me to the
summer school. This work was supported in part by Korea Science and
Engineering Foundation through SRC program.

\end{document}